\documentclass[aps,pre,superscriptaddress,twocolumn,balancelastpage]{revtex4-1}

\usepackage{times}
\usepackage{amsmath,amssymb}
\usepackage{graphicx}
\graphicspath{{figures/}}

\usepackage{verbatim}
\usepackage{color}

\usepackage{flafter}     

\usepackage[colorlinks,bookmarks=false,citecolor=blue,linkcolor=blue,urlcolor=blue]{hyperref}
\usepackage{color}
\usepackage[all]{hypcap} 

\newcommand{\ue}{\text{e}}
\newcommand{\ui}{\text{i}}

\newcommand{\Dq}{D_q}
\newcommand{\Dqinf}{D_q^{\infty}}

\newcommand{\DqN}{\Dq(N)}
\newcommand{\DqjN}{\Dq(j, N)}

\newcommand{\la}{\left\langle}
\newcommand{\ra}{\right\rangle}

\newcommand{\DqNavg}{\overline{\Dq(N)}}
\newcommand{\DqavgCUEN}{\overline{\Dq^{\text{CUE}}(N)}}
\newcommand{\DqavgCOEN}{\overline{\Dq^{\text{COE}}(N)}}

\newcommand{\DOneavgCUEN}{\widetilde{D}_1^{\text{CUE}}(N)}
\newcommand{\DOneavgCOEN}{\widetilde{D}_1^{\text{COE}}(N)}

\newcommand{\IqavgN}{\overline{I_q(N)}}
\newcommand{\IqavgCUEN}{\overline{I_q^{\text{CUE}}(N)}}
\newcommand{\IqavgCOEN}{\overline{I_q^{\text{COE}}(N)}}

\newcommand{\widetildeDq}{\widetilde{D}_q}
\renewcommand{\DqNavg}{\widetildeDq(N)}
\renewcommand{\DqavgCUEN}{\widetildeDq^{\text{CUE}}(N)}
\renewcommand{\DqavgCOEN}{\widetildeDq^{\text{COE}}(N)}

\renewcommand{\IqavgN}{\widetilde{I}_q(N)}
\renewcommand{\IqavgCUEN}{\widetilde{I}_q^{\text{CUE}}(N)}
\renewcommand{\IqavgCOEN}{\widetilde{I}_q^{\text{COE}}(N)}

\newcommand{\DqRMTN}{\Dq^{\text{RMT}}(N)}

\newcommand{\Nup}{n_{\uparrow}}

\newcommand{\ud}{\text{d}}

\makeatletter
\let\Hy@backout\@gobble
\makeatother

\begin{document}

\title{Multifractal dimensions for random matrices, chaotic quantum maps,
and many-body systems}

\author{Arnd B\"acker}
\affiliation{Max-Planck-Institut f\"ur Physik komplexer Systeme,
N\"othnitzer Stra\ss e 38, 01187 Dresden, Germany}
\affiliation{Technische Universit\"at Dresden,
 Institut f\"ur Theoretische Physik and Center for Dynamics,
 01062 Dresden, Germany}

\author{Masudul Haque}
\affiliation{Max-Planck-Institut f\"ur Physik komplexer Systeme,
 N\"othnitzer Stra\ss e 38, 01187  Dresden, Germany}
\affiliation{Department of Theoretical Physics,
Maynooth University, Co. Kildare, Ireland}

\author{Ivan Khaymovich}
\affiliation{Max-Planck-Institut f\"ur Physik komplexer Systeme,
N\"othnitzer Stra\ss e 38, 01187 Dresden, Germany}

\date{\today}

\begin{abstract}

Multifractal dimensions allow for characterizing the localization properties
of states in complex quantum systems.  For ergodic states the finite-size
versions of fractal dimensions converge to unity in the limit of large system
size.  However, the approach to the limiting behavior is remarkably slow.
Thus, an understanding of the scaling and finite-size properties of fractal
dimensions is essential.  We present such a study for random matrix
ensembles, and compare with two chaotic quantum systems --- the kicked rotor
and a spin chain.  For random matrix ensembles we analytically obtain the
finite-size dependence of the mean behavior of the multifractal dimensions,
which provides a lower bound to the typical (logarithmic) averages.
We show that finite statistics has remarkably strong effects, so that even
random matrix computations deviate from analytic results (and show strong
sample-to-sample variation), such that restoring agreement
requires exponentially large sample sizes.
For the quantized standard map (kicked rotor) the multifractal dimensions are
found to follow the random matrix predictions closely, with the same finite
statistics effects.
For a XXZ spin-chain we find significant deviations from the random matrix
prediction~---~the large-size scaling follows a system-specific path towards
unity.  This suggests that local many-body Hamiltonians are ``weakly
ergodic'', in the sense that their eigenfunction statistics deviate from
random matrix theory.

\end{abstract}

\maketitle

\section{Introduction}

Energy eigenstates are integral to the formulation of quantum mechanics.
Except for especially simple systems, eigenstates usually are complicated
objects, described in any basis by a large number of coefficients.  Thus, it is
natural to analyze eigenstate coefficients \emph{statistically}.  Statistical
properties of eigenstates were investigated already very early in the context
of transition strengths for complex nuclei~\cite{BroFloFreMelPanWon1981} which
can be described by random matrix ensembles~\cite{Meh2004}, and are central to
the study of quantum chaos, e.g., in quantum
billiards~\cite{Ber1977b,ShaGoe1984,McDKau1988,AurSte1991,AurSte1993,
LiRob1994,SimEck1996,Pro1997b,Bae2003} and in quantum
maps~\cite{Izr1987, Non1997,KusMosHaa1988, NonVor1998,Bae2003,Non2010}.
They also play a crucial role in
characterizing critical behaviors of Anderson transitions between localized and
metallic phases in disordered systems~\cite{EveMir2008}.  Moreover, the
properties of energy eigenstates are of particular importance for describing
the behavior of isolated quantum many-body systems, e.g., concerning
thermalization~\cite{Deu1991, Sre1994, RigDunOls2008, PolSenSilVen2011,
  AleKafPolRig2016,BorIzrSanZel2016} and many-body
localization~\cite{BasAleAlt2006,OgaHus2007,NanHus2015,AltVos2015,
Imb2016,SerPapAba2017,LuiBar2017b,AleLaf2018}.

The statistical properties of eigenstates have been characterized and studied
in multiple ways.  The distributions of eigenstates have been
examined directly, e.g, for quantum billiards~\cite{ShaGoe1984, McDKau1988,
  AurSte1991, AurSte1993, LiRob1994,SimEck1996, Pro1997b, BaeSch2002a, Bae2007,
  BeuBaeMoeHaq2018,SamJai2018},
for many-body systems~\cite{LucSca2013, BeuBaeMoeHaq2018},
and for quantum maps \cite{Izr1987, Non1997,KusMosHaa1988, NonVor1998,
Bae2003,Non2010} and
random-matrix ensembles~\cite{BogSie2018a, BogSie2018b,TruOss2018}.
The maxima of random waves and chaotic eigenstates have also been
considered~\cite{NonVor1998,AurBaeSchTag1999}.
Eigenstate statistics have often been characterized through
the inverse participation ratio, extensively over several decades
for single-particle systems~\cite{EdwTho1972, Weg1980, KraKin1993, EveMir2000,
  EveMir2008}
and more recently also for many-body systems~\cite{SanRig2010a, LucSca2013,
  BeuAndHaq2015, LuiLafAle2015, TorSan2015, BorIzrSanZel2016, MisPasLuc2016,
  Tsu2017, BeuBaeMoeHaq2018}.
Generalizing the inverse participation ratio, eigenstate statistics has also
been studied through the Shannon and R\'enyi entropies~\cite{JiaSubGruCha2008,
  GirMarGeo2009,SteFurMisPas2009, SteMisPas2010, SanRig2010a, SanRig2010b,
  SteMisPas2011, LuiAleLaf2014}.
Closely related to the R\'enyi entropies are the so-called fractal dimensions
\cite{HenPro1983, HalJenKadProShr1986}, which are the topic of this work.
Analysis of fractal dimensions (`multifractal analysis') is a standard tool in
the study of (single-particle) Anderson localization~\cite{EveMir2008} and has
also been recently applied to eigenstates of many-body quantum
systems~\cite{AtaBog2012, LuiAleLaf2014, SerPapAba2017, BeuBaeMoeHaq2018,
  MacAleLaf2018:p}.

If the $q$-th moment of the eigenstate coefficients scales like
$N^{-(q-1)\Dqinf}$ as a function of the Hilbert space dimension $N$, then the
quantity $\Dqinf\geq 0$ gives the (multi)fractal (Hausdorff) dimension of the
corresponding support set in the limit $N\to\infty$.
The fractal dimensions are particularly useful for distinguishing between
localized and ergodic phases for single-particle lattice systems with disorder.
The Anderson-localized phase is characterized by zero fractal dimensions
$
\Dqinf=0$ for $q>0$, as each eigenstate is localized at a finite number of
sites.  In contrast, so-called ergodic quantum
eigenstates~\cite{EveMir2008} are those states for which at least a
finite fraction of the coefficients in the given basis contribute
significantly, and thus $\Dqinf=1$.

An important class of quantum systems are those
with a well-defined classical limit showing chaotic dynamics
in the sense that one has sensitive dependence on the initial conditions
(positive Lyapunov exponents almost everywhere) and ergodicity
(temporal averages of observables correspond to spatial averages for almost
all initial conditions).
In such cases one expects that the statistical properties of
spectra can be described by those of corresponding
random matrix ensembles~\cite{BohGiaSch1984,GuhMueWei1998,BGSScholarpedia}.
In contrast, many-body systems usually do not have such a classical limit.  We
can define a many-body system as being ``ergodic'' or ``chaotic'' if the
spectral statistics or eigenfunction statistics follow those of one of the
random matrix ensembles.  In either of these cases one expects in the
large-size limit that the fractal dimensions of most eigenstates are equal to
$\Dqinf=1$ for all $q\geq0$.

The fractal dimensions are of particular interest in characterizing
multifractality, in which case
$\Dqinf$ has a nontrivial
$q$-dependence, in contrast to ergodic (localized) states for which
$\Dqinf$ is equal to $1$ ($0$) for all $q\geq0$.  Multifractal
statistics appears at the Anderson localization transition for single-particle
lattice systems~\cite{EveMir2008, BogGir2011, BogGir2012, MenAlcVar2014,
DubGarGeoGirLemMar2014,  UjfVar2015, DubGarGeoGirLemMar2015,LinRod2017a}.
In addition, recent examples have reported (multi)fractal phases
extending over a whole range of parameters~\cite{LucAltKraSca2014,
  AltCueIofKra2016, KraAltIof2018,
  KraKhaCueAmi2015, PinKraAltIof2017, RoyKhaDasMoe2018, BerTomKhaSca2018,
  TomAmiBerKhaKra2019,NosKhaKra2019, SmeKecBoiIsaNevAlt2018:p,
  KecSmeMcCDenMohIsaBoiAltNev2018:p, FaoFeiIof2018:p, MicMonAlt2019:p,
  PinTabSer2019:p, NosKha2019:p}.
Multifractal wavefunctions have been found for some quantum maps
\cite{MarGirGeo2008, MarGarGirGeo2010, DubGarGeoGirLemMar2014,
  DubGarGeoGirLemMar2015}.
For local many-body quantum Hamiltonians, the ground states have been found to
display multifractal behavior, even in cases for which eigenstates at the
center of the many-body spectrum show random-matrix behavior~\cite{AtaBog2012,
  AtaBog2014b, LuiAleLaf2014,BeuBaeMoeHaq2018, LinRod2017b, LinBucRod2019}.
Also, the question of the existence of a multifractal phase in the vicinity of
the many-body localization transition as well as its relation to the slow
dynamical phases is under active debate~\cite{LuiAleLaf2014,
  MacAleLaf2018:p,LezBerBar2019,AltCueIofKra2016, KraAltIof2018,
  TikMir2016,SonTikMir2017,TikMirSkv2016,BirTar2017,AleLaf2018}).

In this paper, we examine the finite-size dependence of fractal dimensions
($N$-dependence of $\DqN$) for eigenstates of random matrices and of nominally
chaotic systems.  The eigenstates of these systems are expected to be at least
weakly ergodic.  Ergodic states are considered to be less exotic than
multifractal states, since the large-$N$ limit is simple.  However, we will
present highly nontrivial scaling behaviors: $\DqN$ approaches unity extremely
slowly and with large eigenstate-to-eigenstate fluctuations.  We will first
present analytical and numerical results for the case of random-matrix
ensembles, namely the circular orthogonal (COE) and unitary (CUE) ensembles.
These results will then be compared to two physical systems which are expected
to have ergodic behavior.  The first is a paradigmatic model from quantum
chaos: the quantum kicked rotor whose corresponding classical dynamics is given
by the standard map.  We will show that the multifractal properties of the
quantized standard map with strongly chaotic classical dynamics follow the CUE
predictions very closely.  We then consider a non-integrable quantum spin
chain. In this case the comparison is substantially more subtle because only
the center of the many-body spectrum (``infinite-temperature'' states) is
expected to behave ergodically.  We present numerical evidence that the
behavior of many-body eigenstates is only ``weakly ergodic'', in the sense that
$\DqN$ approaches unity for $N\to\infty$ but follows a different
system-specific path compared to the COE case.

The paper is structured as follows.  We introduce the fractal dimensions in
Section~\ref{Sec:defs}, in particular the mean and typical averages.  In
Section~\ref{Sec:RMT} we present analytic derivations for the random matrix
ensembles and compare with numerical calculations for COE and CUE ensembles.
In Section~\ref{Sec:quantum_maps} we present calculations of $\DqN$ for the
chaotic quantum map and also compare with random matrix results
Section~\ref{Sec:manybody} treats as example of a many-body quantum system
a spin chain in the chaotic regime, and an analysis of $\DqN$ is presented.
In Section~\ref{Sec:summary} we summarize and point out open questions.

\section{Fractal dimensions \label{Sec:defs}}

To characterize the properties of a given state $|\Psi_j\rangle$
consider its expansion coefficients $c_i^{(j)}$
in some (finite) orthonormal basis $\{|\psi_i \rangle\}$, i.e.\
$|\Psi_j\rangle = \sum_{i=1}^N c_i^{(j)} |\psi_i \rangle$.
Based on the moments
\begin{equation}\label{eq:I_q}
   I_q(j, N) = \sum_{i=1}^N |c_i^{(j)}|^{2q}
\end{equation}
one defines the (finite-$N$) fractal dimensions for
the given state
\begin{align}
  \DqjN &= - \frac{1}{q-1} \frac{1}{\ln N} \ln  I_q(j, N)\\
        &= - \frac{1}{q-1} \frac{1}{\ln N}
             \ln  \left(\sum_{i=1}^N |c_i^{(j)}|^{2q}\right) .
\end{align}
For fixed $N$ the fractal dimensions are monotonically
decreasing functions of $q$ with $0\leq\DqN\leq1$ for $q\ge 0$.
In the limit $q\to 1$ one gets by l'H{\^o}pital's rule
the Shannon information dimension
\begin{equation} \label{eq:D-1-N-j-def}
  D_1(j, N)
      = - \frac{1}{\ln(N)} \sum_{i=1}^N  |c_i^{(j)}|^{2} \, \ln |c_i^{(j)}|^{2}.
\end{equation}
One may now consider an average
over an ensemble of states, which is denoted by
\begin{equation} \label{eq:DqN}
  \DqN = \la \DqjN\ra
       \equiv -\frac{1}{q-1}\frac{1}{\ln N} \la \ln I_q(j,N)\ra .
\end{equation}
Finally, the fractal dimensions $\Dqinf$ are defined
in the limit $N\to\infty$~\cite{MirEve2000}, i.e.\
\begin{equation}
  \Dqinf \equiv \lim_{N\to\infty} \DqN.
\end{equation}
If $\Dqinf$ depends on $q>0$ in a nontrivial way,
the states are multifractal.
For constant $\Dqinf <1$ the states are fractal,
ergodic behavior corresponds to $\Dqinf=1$,
and localized states correspond to $\Dqinf = 0$.

Thus the fractal dimensions $\Dqinf$ describe the asymptotic
scaling behavior of the moments of typical eigenstates as $N\to \infty$, i.e.\
\begin{equation} \label{eq:I-q-typ}
\begin{split}
  \la I_q(j, N) \ra_{\text{typ}}  &\equiv
   \exp\left[\la \ln I_q(j,N)\ra\right]\\
   &\stackrel{N\to\infty}\sim N^{-\Dqinf (q-1)}.
\end{split}
\end{equation}
Numerically, $\Dqinf$ can only be estimated by extrapolating the results of
finite-$N$ computations using Eq.~\eqref{eq:DqN}.

The leading size-dependence of $\DqN$ is often of the form
\begin{equation}
  \DqN\sim \Dqinf - f_q/\ln{N} ,
\end{equation}
so that using Eq.~\eqref{eq:DqN} the moments can be written as
\begin{equation} \label{eq:IqEqcqNDq}
 \la I_q(j, N) \ra_{\text{typ}} = N^{-\DqN (q-1)} \simeq c_q N^{-\Dqinf (q-1)},
\end{equation}
with $c_q=\ue^{(q-1)f_q}$.  When the finite-size correction to $\Dqinf$
is not exactly or solely of the form proportional to
$1/\ln{N}$, the pre-factor $c_q$ acquires a weak dependence on $N$.

Random matrix theory allows for a universal description of the statistical
properties of ergodic eigenstates in many different situations.  Thus it should
also provide a prediction for the finite-$N$ scaling of $\DqN$, where the
average in Eq.~\eqref{eq:DqN} is performed over a suitable random matrix
ensemble.
However, an analytical computation of the ensemble average in
Eq.~\eqref{eq:DqN} over the logarithm of the moments is a daunting task.  Thus
instead we will use the ensemble averaged moments
$\IqavgN\equiv \la I_q(j, N) \ra$ and take the logarithm afterwards, i.e.\
\begin{equation} \label{eq:D-q-N-avg}
  \DqNavg = - \frac{1}{q-1} \frac{1}{\ln N} \ln  \IqavgN .
\end{equation}
By Jensen's inequality
$\DqNavg$ provides a lower bound to $\DqN$,
\begin{equation} \label{eq:Jensen}
   \DqNavg  \le \DqN
\end{equation}
as the logarithm is a concave function.
In particular $\DqNavg \to 1$ implies $\DqN \to 1$.

\section{Random matrix predictions}\label{Sec:RMT}

As specific random matrix ensembles we consider
the circular unitary ensemble (CUE)
of complex unitary matrices, describing
systems without any antiunitary symmetries
and the circular orthogonal ensemble (COE)
of real orthogonal matrices, describing
systems with one antiunitary symmetry, e.g.\ time-reversal.
Note that the results for the
eigenvector statistics of the CUE and COE also apply to the Gaussian unitary
ensemble (GUE) and the Gaussian orthogonal ensemble (GOE), respectively.

\subsection{Circular orthogonal ensemble} \label{sec:COE}

\begin{figure}[b]
  \includegraphics{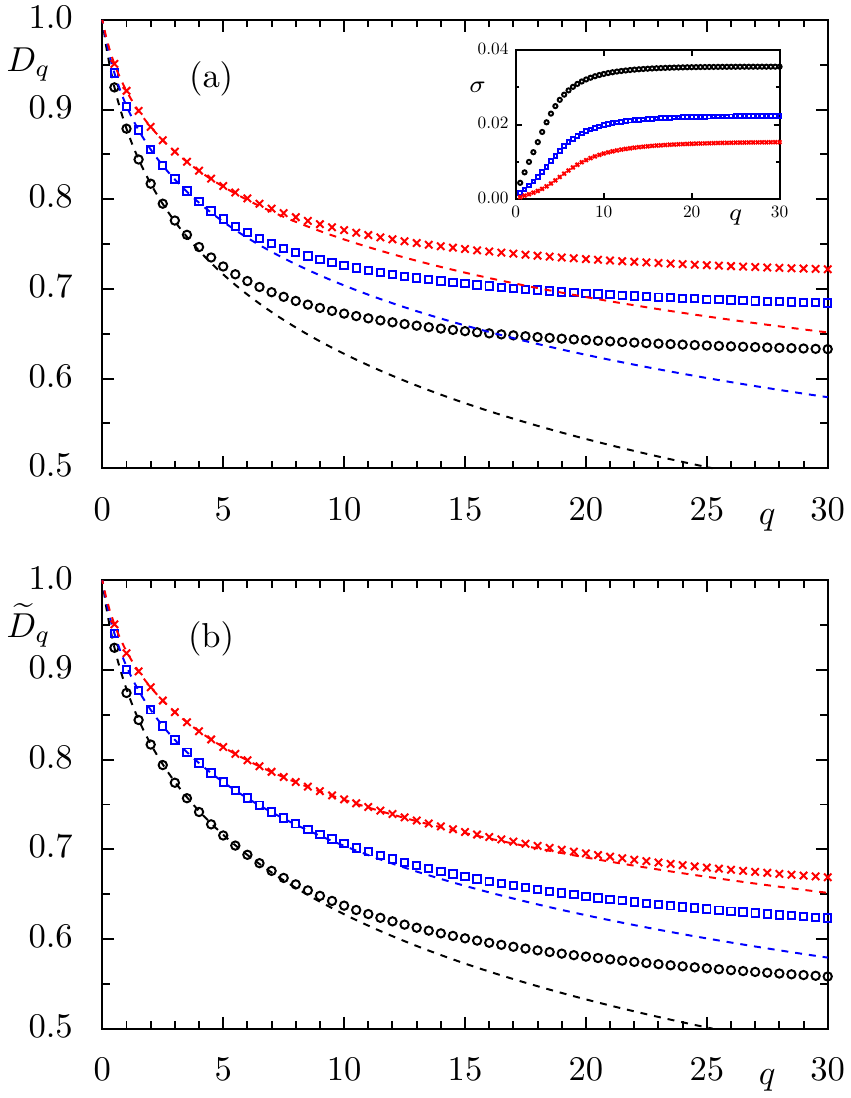}
  \caption{Fractal dimensions (a) $\DqN$ of typical and (b) $\DqNavg$
           of mean eigenstate moments
           for the COE for $N=400, 2000, 10000$
           (black circles, blue squares, red crosses)
           in comparison with
           $\DqavgCOEN$, Eq.~\eqref{eq:D-q-COE-typical}, dashed lines.
           The inset in (a) shows the standard deviation
           $\sigma(q)$ of the fluctuations of $\DqjN$ around $\DqN$.
           }
  \label{fig:D-q-COE-N}
\end{figure}

For the COE the eigenvectors can be chosen to be real
and the only requirement for the coefficients $c_i^{(j)}$
is the normalization
\begin{equation} \label{eq:COE-normalization}
  \sum_{i=1}^N \left(c_i^{(j)}\right)^2 = 1 .
\end{equation}
This condition implies
that the probability density of one (rescaled) component
$N \bigl(c_i^{(j)}\bigr)^2$ to have a specific value $\eta$
is given by~\cite[Eq.~(7.5)]{BroFloFreMelPanWon1981}
\begin{equation} \label{eq:P-N-COE}
  P_N^{\text{COE}}(\eta)
     = \frac{1}{\sqrt{\pi N \eta}} \frac{\Gamma(N/2)}{\Gamma( (N-1)/2)}
       (1-\eta/N)^{(N-3)/2} .
\end{equation}
The corresponding eigenfunction moments~\eqref{eq:I_q} calculated
from this distribution are
\begin{align}
  \IqavgCOEN
       & =  \frac{N}{\sqrt{\pi}} \frac{\Gamma(N/2) \Gamma(q+1/2)}
                                      {\Gamma(q+N/2)}
         . \label{eq:I-q-N-COE}
\end{align}
For large $N$ one gets
\begin{align}
\!\!\!
  \IqavgCOEN
       & \simeq  \frac{\Gamma(q+1/2)(2\ue)^q}{\sqrt{\pi}} N^{-(q-1)}
           \\
       & \;\;\;\;
          \times \left(1-\frac{2}{N}\right)^{\frac{N-1}2}
                 \left(1-\frac{2(q-1)}{N}\right)^{q+\frac{N-1}2}\\
       & \simeq  \frac{\Gamma(q+1/2)2^q}{\sqrt{\pi}} N^{-(q-1)} + O(N^{-q})\\
       &  \sim N^{-(q-1)}.
\end{align}

Note that one obtains from~\eqref{eq:P-N-COE}
in the limit of large $N$ the so-called
Porter-Thomas distribution~\cite{PorTho1956}
\begin{equation} \label{eq:distribution-N-limit-COE}
  P^{\text{COE}}(\eta)= \frac{1}{\sqrt{2\pi  \eta }} \exp(-\eta/2) .
\end{equation}

Based on the moments~\eqref{eq:I-q-N-COE}, inserted
in Eq.~\eqref{eq:D-q-N-avg}, one gets the COE prediction
for the finite-$N$ scaling of the fractal dimensions
\begin{align}
 \DqavgCOEN
    & = - \frac{1}{(q-1) \ln N}
           \ln \left( \IqavgCOEN \right) \\
    & = - \frac{1}{(q-1) \ln N}
           \ln \left( \frac{N \Gamma(N/2) \Gamma(q+1/2)}
                           {\sqrt{\pi}\Gamma(q+N/2)}  \right)
           \label{eq:D-q-COE-typical} \\
      &\simeq 1 - \frac{1}{(q-1)\ln N}
           \ln \left(\frac{\Gamma(q+1/2)2^q}{\sqrt{\pi}} \right).
\end{align}
For $q=1$ this gives $\DOneavgCOEN = 1 - (\ln 2 + \psi(3/2))/\ln N$,
with the digamma function $\psi(x)=\Gamma'(x)/\Gamma(x)$~\cite[5.2.E2]{DLMF}.
The fractal dimensions approach
$\DqavgCOEN\to 1$
with  logarithmic corrections $\sim 1/\ln N$.
For the pre-factor $c_q$ introduced in Eq.~\eqref{eq:IqEqcqNDq}, this gives
\begin{equation} \label{eq:c-q-COE}
  c_q^{\text{COE}} \simeq \Gamma(q+1/2)2^q/\sqrt{\pi}.
\end{equation}

Figure~\ref{fig:D-q-COE-N}(a) shows $\DqN$ for $N=400$, $2000$, and $10000$,
each computed from one realization of the COE, numerically generated as
described in~\cite{Mez2007}.  The curves are still very far from $\Dq = 1$, but
a slow logarithmic approach with increasing $N$ is clearly seen.  The
analytical result $\DqavgCOEN$, Eq.~\eqref{eq:D-q-COE-typical}, provides
according to the inequality~\eqref{eq:Jensen}, a lower bound.  This bound even
gives a good approximation up to some value of $q$, which increases with
increasing $N$.
In Fig.~\ref{fig:D-q-COE-N}(b) we
show $\DqNavg$ for the COE, i.e.\
for one realization the moments $\IqavgCOEN$
are computed and then Eq.~\eqref{eq:D-q-N-avg} is used.
The agreement with  the analytical result $\DqavgCOEN$,
Eq.~\eqref{eq:D-q-COE-typical}, is much better.
However, for larger values of $q$, there are still prominent deviations from
the analytic predictions.  We will discuss the origin of these deviations in
Sec.~\ref{sec:finite-N-statistics}.

The inset in Fig.~\ref{fig:D-q-COE-N}(a) shows
the standard deviation
$\sigma(q)$ of the fluctuations of $\DqjN$ around $\DqN$.
For fixed $N$, the state-to-state fluctuations increase with
increasing $q$, and appear to eventually saturate.
Larger values of $N$ lead to smaller fluctuations.

\subsection{Circular unitary ensemble}

\begin{figure}[b]
  \includegraphics{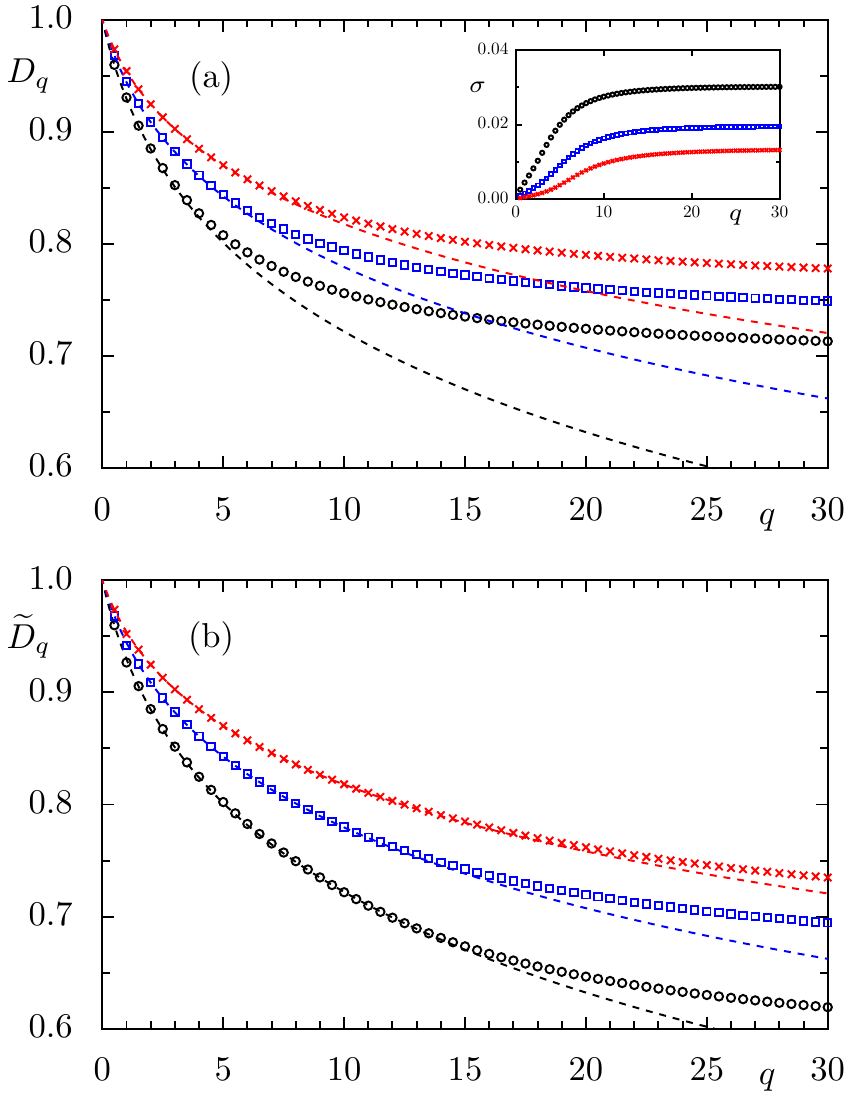}
  \caption{Fractal dimensions (a) $\DqN$ of typical and
           (b) $\DqNavg$ of mean eigenstate moments
           for the CUE for $N=400, 2000, 10000$
           (black circles, blue squares, red crosses)
           in comparison with
           $\DqavgCUEN$, Eq.~\eqref{eq:D-q-CUE-typical}, dashed lines.
           The inset in (a) shows the standard deviation
           $\sigma(q)$ of the fluctuations of $\DqjN$ around $\DqN$.
           }
  \label{fig:D-q-CUE-N}
\end{figure}

For the CUE the eigenvectors are complex,
fulfilling the normalization condition
\begin{equation} \label{eq:CUE-normalization}
  \sum_{i=1}^N \left|c_i^{(j)}\right|^2 = 1 .
\end{equation}
This implies that the probability density of one (rescaled) component
$N |c_i^{(j)}|^2$ to have a specific value $\eta$
is given by~\cite{KusMosHaa1988}
\begin{equation} \label{eq:P-N-CUE}
  P_N^{\text{CUE}}(\eta)
     = (1-1/N) \left(1-\eta/N\right)^{N-2}.
\end{equation}
The corresponding eigenfunction moments~\eqref{eq:I_q}
calculated from this distribution are
\begin{align}
  \IqavgCUEN
       & =  \frac{q! N!}{(N+q-1)!}
         \label{eq:I-q-N-CUE}.
\end{align}
For large $N$ one gets
\begin{align}
\!\!\!
  \IqavgCUEN
       & \simeq  q!\ue^q N^{-(q-1)}
                 \\
       & \;\;\;\;
             \times    \left(1-\frac{1}{N}\right)^{N-\frac1 2}
                       \left(1-\frac{q-1}{N}\right)^{q+N-\frac1 2}\\
       & \simeq  q! N^{-(q-1)} + O(N^{-q})\\
       & \sim N^{-(q-1)}.
\end{align}
Note that  one obtains from Eq.~\eqref{eq:P-N-CUE},
in the limit of large $N$,
\begin{equation} \label{eq:distribution-N-limit-CUE}
   P^{\text{CUE}}(\eta)  = \exp(-\eta) \;\;.
\end{equation}

Based on the moments~\eqref{eq:I-q-N-CUE},
inserted in Eq.~\eqref{eq:D-q-N-avg}, one gets the CUE prediction
for the $N$-dependence of the fractal dimensions
\begin{align}
  \DqavgCUEN
    & = - \frac{1}{(q-1) \ln N}
           \ln \left( N \IqavgCUEN \right) \\
        &= - \frac{1}{(q-1) \ln N}
           \ln \left( \frac{q! N!}{(N-1+q)!} \right)
           \label{eq:D-q-CUE-typical} \\
        & \simeq 1 - \frac{\ln (q!)}{(q-1)\ln N} .
\end{align}
For $q=1$ this gives $\DOneavgCUEN = 1-(1-\gamma)/\ln N$, where
$\gamma\simeq0.577216$ is Euler's constant.
Clearly, as $N\to \infty$ the fractal dimension
approaches $\DqavgCUEN \to 1$ with
logarithmic corrections $\sim 1/\ln N$, as in the COE case.
For the pre-factor $c_q$ introduced in Eq.~\eqref{eq:IqEqcqNDq},
this gives
\begin{equation}
  c_q^{\text{CUE}} \simeq q!.
\end{equation}

\begin{figure}
  \includegraphics{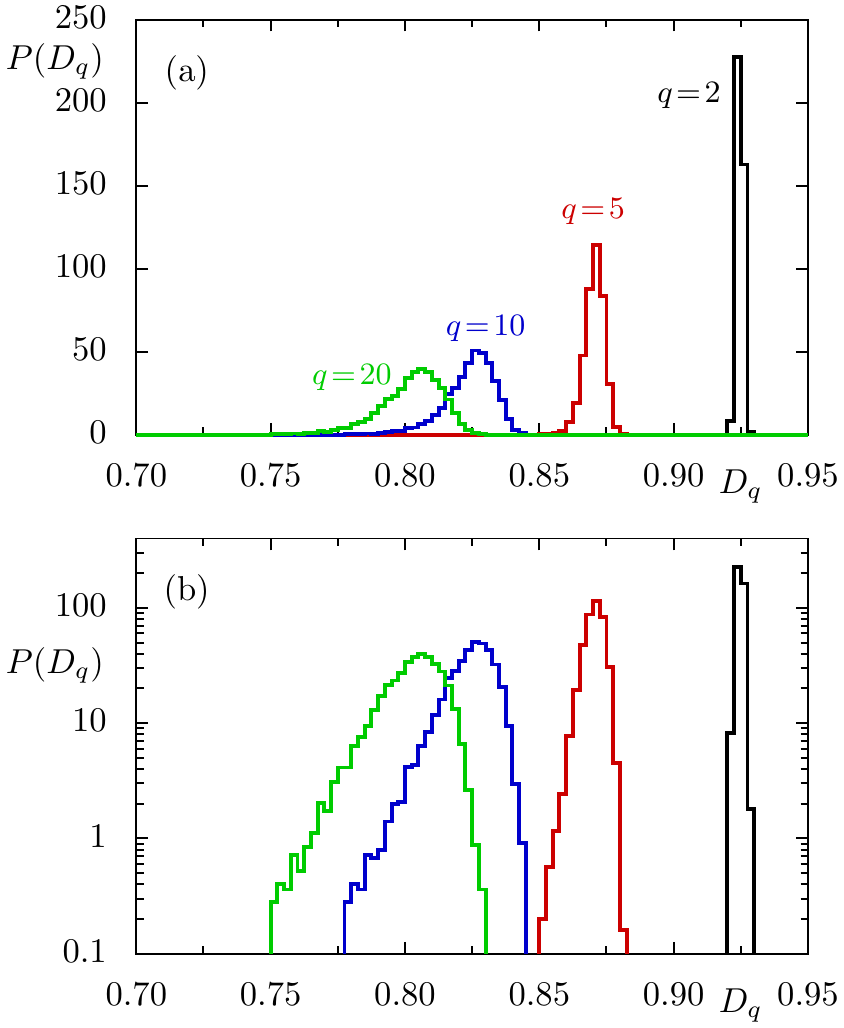}
  \caption{(a) Distributions $P(\DqjN)$ for one realization of the
           CUE for $N=10000$ and $q=2, 5, 10, 20$.
           (b) Semi-logarithmic representation.}
  \label{fig:D-q-CUE-distrib}
\end{figure}

Figure~\ref{fig:D-q-CUE-N}(a) shows a comparison
of $\DqN$ for the CUE
with the lower-bound $\DqavgCUEN$, Eq.~\eqref{eq:D-q-CUE-typical}
for $N=400$, $2000$ and $10000$, each using one realization.
Similarly to the case of the COE
one finds an overall slow logarithmic approach towards
$\Dq = 1$ with increasing $N$.
The deviations from
the lower bound $\DqavgCUEN$, Eq.~\eqref{eq:D-q-CUE-typical}
start for larger $q$ with increasing $N$ than for the COE.
Also for $\DqNavg$ for the CUE,
shown in Fig.~\ref{fig:D-q-CUE-N}(b),
the agreement with  the analytical result $\DqavgCUEN$,
Eq.~\eqref{eq:D-q-CUE-typical} is better,
however, again with unexpected prominent deviations for larger $q$.

In Figs.~\ref{fig:D-q-COE-N}(a) and \ref{fig:D-q-CUE-N}(a), we have presented
$\DqN$, which is the average of $\DqjN$ over $N$ states.  However, there is
quite a variation in the values of $\DqjN$ themselves.  This is already
indicated by the standard deviation $\sigma(q)$, as shown in the insets
Fig.~\ref{fig:D-q-COE-N}(a) and Fig.~\ref{fig:D-q-CUE-N}(a) but better seen in
the full distribution of $\DqjN$.  Figure~\ref{fig:D-q-CUE-distrib}(a) shows
the histograms for $\DqjN$ for the CUE for $N=10000$ and $q=2, 5, 10, 20$.
With increasing $q$, the mean decreases while the variance increases.  This can
also be understood intuitively, as larger values of $q$ correspond to higher
moments of the eigenstate coefficients which therefore emphasizes the tails of
the distribution of the coefficients.  The semi-logarithmic representation in
Fig.~\ref{fig:D-q-CUE-distrib}(b) shows that the tails towards smaller $\Dq$
become approximately a straight line, i.e.\ show exponential behavior, while
the tails towards larger $\Dq$ are close to a Gaussian decay.

Based on the properties of the distributions $P(\DqjN)$
one can draw several conclusions about the behavior of $\DqN$ and $\DqNavg$:
(i) The fact that $P(\DqjN)$ has a rapid decay in both directions ensures
that the values of $\DqN$ and $\DqNavg$ have similar orders of magnitude,
as observed in Figs.~\ref{fig:D-q-COE-N} and~\ref{fig:D-q-CUE-N}.
(ii) However, for larger $q$ the numerically computed $\DqNavg$
are always above the  analytical results $\DqavgCOEN$ and $\DqavgCUEN$,
see Figs.~\ref{fig:D-q-COE-N}(b) and~\ref{fig:D-q-CUE-N}(b).
The origin for this is the skewness of the distributions  $P(\DqjN)$
towards lower values of $\DqjN$ together with the monotonic decay
of the function $N^{-(q-1)\Dq}$ with $\Dq$ at $q>1$.
Indeed, the analytical results $\DqavgCOEN$ and $\DqavgCUEN$
can also be obtained by the integral
\begin{equation}\label{eq:P_Dq_avg}
N^{-(q-1)\DqNavg} = \int P(\Dq) N^{-(q-1)\Dq}d\Dq \ ,
\end{equation}
where the integrand is more skewed to the left in comparison with $P(\Dq)$,
but still decays rapidly.
The numerically sampled $\DqNavg$ is governed by the most probable
values around the maximum of the integrand in~\eqref{eq:P_Dq_avg}.
Thus it deviates to larger values from $\DqavgCOEN$.
(iii) The shape of the distribution of $P(D_q)$
appears to stabilize with increasing $q$; see for example
the $q=10$ and $q=20$ distributions  in Fig.~\ref{fig:D-q-CUE-distrib}.
This corresponds to the saturation of the standard deviation $\sigma(q)$
at large $q$, seen in the insets of
 Figs.~\ref{fig:D-q-COE-N}(a) and \ref{fig:D-q-CUE-N}(a)

\subsection{Finite-statistics corrections}
\label{sec:finite-N-statistics}

In this section we consider corrections to the moments $\IqavgN$ and fractal
dimensions $\DqN$ and $\DqNavg$ due to finite statistics.  This allows to
estimate the value of $q$ above which the numerical calculations deviate from
analytic predictions.

We consider a situation where one obtains the data from a finite number $N_r$
of eigenstates, which may be from one or several (e.g.\ disorder) realizations.
Statistical errors come into play because of the finiteness of $N_r$.
We can characterize these errors by considering how the distribution
$P(N|c_i^{(j)}|^2=\eta)\equiv P(\eta)$ is numerically approximated by a
histogram.  The histogram is normalized by $N_r$ and has bin sizes
$\Delta\eta$.  We first consider the bin sizes $\Delta\eta$ to be independent
of $\eta$.
At the edge of the distribution, i.e.\ for larger values of $\eta$,
the number of counts per bin is smaller,
and hence statistically less reliable.  When there are
only a few counts, $C\sim O(1)$, statistical errors become significant. The bin
at which this occurs, i.e., the value $\eta=\eta^*$, is given by the condition
\begin{gather}\label{eq:N_r_P_Deltaeta_sim_C}
  N_{r} P(\eta^*)\Delta \eta \simeq C \ .
\end{gather}

For the CUE case, using the exponential~\eqref{eq:distribution-N-limit-CUE}
as large-$N$ approximation, one obtains the condition
\begin{equation}
\eta^*(N_r) \simeq
\ln\left[\frac{N_r\Delta\eta}{C}\right] = \ln{\bar{N}_r},
\end{equation}
where $\bar{N}_r \equiv \left(N_r\Delta\eta/C\right)$.

For the COE case, the large-$N$ approximation is the
Porter-Thomas distribution~\eqref{eq:distribution-N-limit-COE}.
For this we cannot solve Eq.~\eqref{eq:N_r_P_Deltaeta_sim_C} for $\eta^*$
in closed form, but approximating iteratively, we obtain
\begin{eqnarray}
  \eta^*(N_r) &\simeq& 2\ln{\bar{N}_r} + \ln\eta \\
  &=& 2\ln{\bar{N}_r} + \ln\left[2\ln{\bar{N}_r} + \ln\eta\right] = \ldots,
\end{eqnarray}
In the iterative solution for the COE case, the corrections to the leading term
are either constant or multiple-logarithmic functions of $\bar{N}_r$; for our
estimate we neglect these weakly varying functions and keep only the leading
($2\ln{\bar{N}_r}$) term.
Thus we get
\begin{gather}\label{eq:eta_star}
  \eta^*(N_r) \simeq \frac{2}{\beta}\ln{\bar{N}_r}
              \simeq \frac{2}{\beta}\ln{N_r},
\end{gather}
up to $O(1)$ constants and additive weaker functions of $N_r$.
Here $\beta=1$ for the COE and $\beta=2$ for the CUE.

\begin{figure}
  \includegraphics{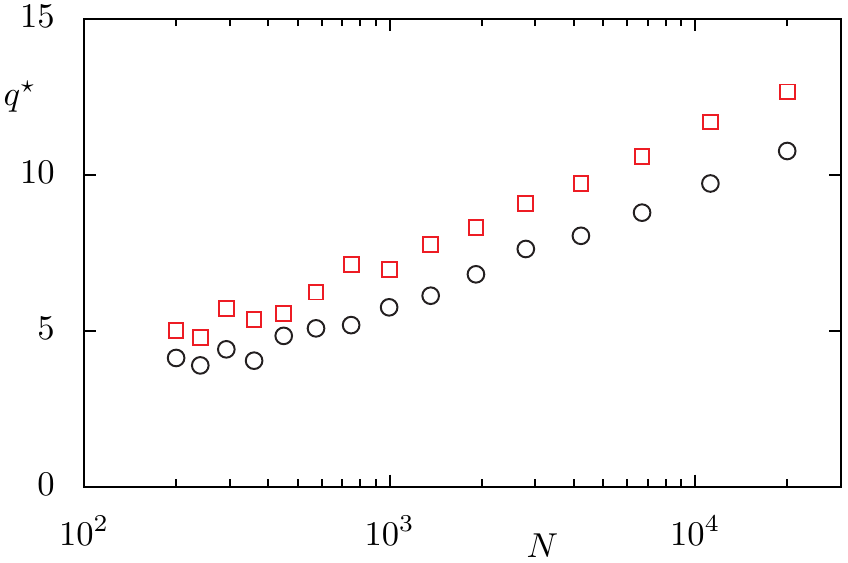}
  \caption{Plot of the moment $q^\star(N)$
           from which on RMT prediction and data beginn to differ
           by more than $f=0.009$;
           COE (black circles) and CUE (red squares).
         }
  \label{fig:q-star}
\end{figure}

Note that there is no fundamental reason for the binning to be linear, i.e.,
for $\Delta\eta$ to be independent of $\eta$.  If one uses logarithmic binning,
$\Delta\eta\propto\eta$, one obtains corrections to the above estimate which
are of double-logarithmic form, and hence can be neglected as done above.

To obtain an estimate for the value $q^*$ of $q$ at which statistical
errors become significant, we have to to relate $q$ to $\eta$.
Writing the $q$-th moment~\eqref{eq:I_q} as
\begin{align}
N^{q-1} \la I_q(j,N) \ra & = \int_0^N \ud\eta\, \eta^q P(\eta) \\
                        & =  \int_0^N \ud\eta\, \ue^{q\ln \eta +\ln P(\eta)} \ ,
\end{align}
we use the saddle point approximation to note that the main contribution comes
from the value of $\eta$ that maximizes the exponent:
\begin{multline}\label{eq:q-saddle-point}
0 = \frac{\ud}{\ud\eta}\left[q\ln \eta +\ln P(\eta)\right]
= \frac{q}{\eta}+ \frac{\ud \ln P(\eta)}{\ud \eta} \\=
\frac{q}{\eta} -\frac{\beta}{2}+\frac{\beta-2}{2\eta}\ .
\end{multline}
Thus the main contribution to the $q$-th moment
comes from $\eta(q) = (2q+\beta-2)/\beta$.  When $q$ gets so large that
this $\eta(q)$ exceeds $\eta^*$, statistical errors become significant.  Thus
the value of $q$ beyond which statistical errors are significant is
\begin{equation}\label{eq:matching}
q^\star(N_r) = \frac{\beta}{2} \eta^\star(N_r) + 1- \frac{\beta}{2}
\simeq \ln N_r .
\end{equation}

This estimate neglects $O(1)$ constants and weaker (double-logarithmic)
dependences on $N_r$.  In addition, the argument relies on some constants that
cannot by nature be firmly specified, such as the bin count $C$ at which we
consider statistical errors to become significant.  Finally, the deviation
between numerical and analytical predictions, seen in Figs.~\ref{fig:D-q-COE-N}
and~\ref{fig:D-q-CUE-N}, gradually increase with $q$ and do not start at a
sharply defined value of $q^*$. For all these reasons, we do not expect the
estimate to be quantitatively accurate.

Figure~\ref{fig:q-star} shows numerical estimates of
$q^\star(N)$ as a function of $\ln N$.  This gives an idea of how well the data
for $\DqN$ and $\DqNavg$ for one realization of the CUE and the COE are
described by $\DqavgCUEN$, Eq.~\eqref{eq:D-q-COE-typical}, and $\DqavgCOEN$,
Eq.~\eqref{eq:D-q-CUE-typical}, respectively.  We determine $q^\star(N)$ as the
lowest value of $q$ for which $\DqN$ or $\DqNavg$ differ from the random matrix
prediction by more than $f=0.009$, i.e.\
\begin{equation}
   |\DqN - \DqRMTN | \le f  \qquad \text{ for $q < q^\star(N)$.}
\end{equation}
Here $f=0.009$ approximately corresponds to the vertical extent of the symbols
in Figs.~\ref{fig:D-q-COE-N} and~\ref{fig:D-q-CUE-N}.  Of course, this estimate
will depend on the choice of $f$, which is arbitrary.  Despite this uncertainty
and those discussed above, an approximate straight-line dependence is observed,
i.e.\
\begin{equation}
  q^\star(N) \sim \ln N,
\end{equation}
in agreement with the theoretical expectation \eqref{eq:matching} for $N_r=N$.

\section{Chaotic quantum map}\label{Sec:quantum_maps}

For quantum systems whose corresponding classical dynamics is fully chaotic one
expects that the statistics of eigenvalues and eigenstates can be described by
random matrix theory.  Still, even if the spectral
statistics, e.g.\ for the level-spacing distribution, follow the corresponding
random matrix results, this need not hold equally well for the statistics of
eigenstates.
Thus we now investigate, starting with a single-particle system,
how well the results for the scaling of the fractal dimensions are
fulfilled for different types of chaotic quantum systems.
In particular deviations may reveal interesting physics.

\begin{figure}[t]
  \includegraphics{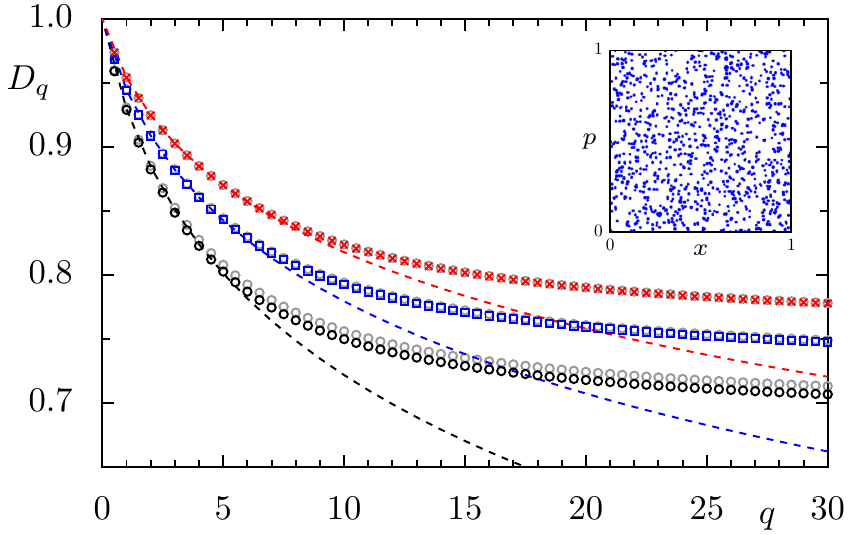}
  \caption{Fractal dimensions $\DqN$
           for the quantized standard map for $N=400, 2000, 10000$
           (circles, squares, crosses)
           in comparison with $\DqavgCUEN$,
           Eq.~\eqref{eq:D-q-CUE-typical}, dashed lines.
           Also shown are the CUE results of Fig.~\ref{fig:D-q-CUE-N}(a)
           as grey circles.
           The inset shows 1000 iterates of
           the standard map~\eqref{eq:standard-map}.
           }
  \label{fig:D-q-std-N}
\end{figure}

As a prototypical example of a system with chaotic classical dynamics
we consider a time-periodically kicked system whose Hamiltonian reads
\begin{equation} \label{eq:hamiltonian}
  H(x, p, t) = \frac{1}{2} p^2 + V(x) \sum_{n=-\infty}^{\infty} \delta(t-n).
\end{equation}
Here the sum describes a periodic sequence of kicks
with unit time as kicking period.
For $V(x) = \frac{K}{4\pi^2} \cos(2\pi x)$ one
obtains the so-called kicked rotor.
Its stroboscopic dynamics considered before consecutive kicks,
gives the area-preserving standard map~\cite{Chi1979},
$(x, p) \mapsto (x', p')$,
\begin{align} \label{eq:standard-map}
     x' &= x + p'\\
     p' &= p + \tfrac{K}{2\pi} \sin(2\pi x) ,
\end{align}
for which we consider $x, p \in[0, 1[$
with periodic boundary conditions so that the phase space
is a two-dimensional torus.
For sufficiently large kicking strength $K$
the standard map is strongly chaotic \cite{Chi1979,Gor2012}.
As example we use $K=9$, see the inset in Fig.~\ref{fig:D-q-std-N},
for which numerically no regular islands on any relevant scales have been found.

Quantum mechanically, the torus phase space
leads to a finite Hilbert space of dimension $N$,
see e.g.\ Refs.~\cite{BerBalTabVor1979, HanBer1980, ChaShi1986, KeaMezRob1999,
  DegGra2003bwcrossref}.
The effective Planck constant is $h=1/N$
and $N\to \infty$ corresponds to the semiclassical limit.
The quantum time evolution between consecutive kicks
is given by a unitary time-evolution operator
which can be represented in position space by a matrix
with elements
\begin{align}    U(n^{\prime},n)
     & =  \frac{1}{N}  \exp\left(- \ui N \frac{K}{2 \pi}
            \cos\left(\frac{2\pi}{N}(n+\alpha)\right)\right) \nonumber \\
     & \;\; \times \;
        \sum_{m=0}^{N-1} \exp\left(-\frac{\pi \ui }{N} (m+\beta)^2\right)
        \label{eq:quantum-map} \\
     & \;\; \quad \quad \times \;
               \exp\left(\frac{2 \pi \ui}{N} (m+\beta)(n-n^{\prime})\right),
       \nonumber
\end{align}
where $n,n'\in\{0,1, ..., N-1\}$.
Thus one gets the eigenvalue problem
\begin{equation}
  U |\psi_j\rangle = \ue^{\ui \varphi_n} |\psi_j\rangle,
\end{equation}
with eigenphases $\varphi_n\in[0, 2\pi[$ as
all eigenvalues lie on the unit circle due to the unitarity of $U$.

The quantum phases $\beta$ and $\alpha$ in Eq.~\eqref{eq:quantum-map} determine
the boundary conditions due to the periodicity in position and momentum,
respectively.  Choosing $(\alpha, \beta) = (0.2, 0.24)$ ensures that both time
reversal symmetry and parity are broken, so that the consecutive level spacing
distribution of this quantized standard map follows the prediction for the CUE.

\begin{figure*}
  \includegraphics{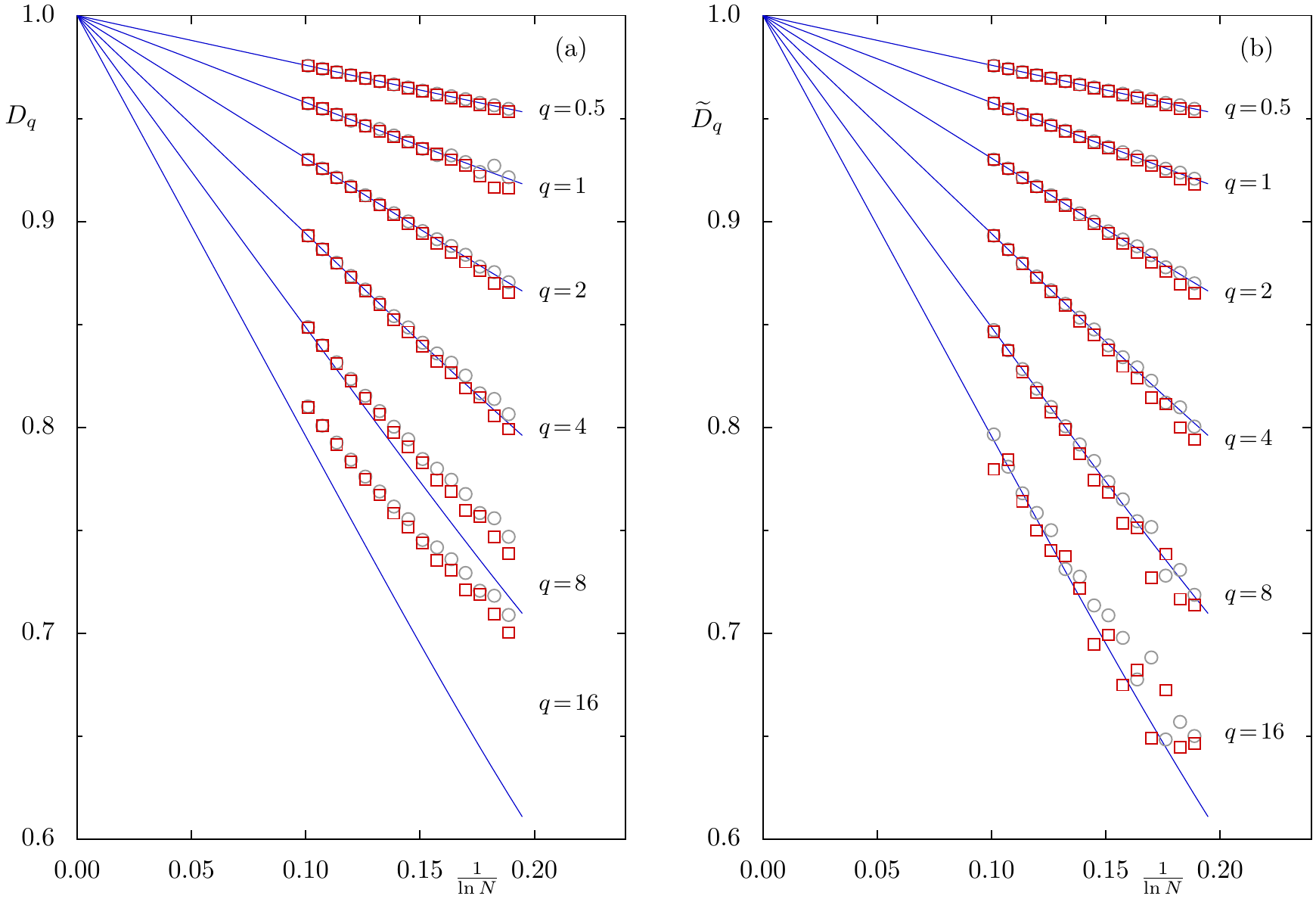}
  \caption{Fractal dimensions (a) $\DqN$ and (b) $\DqNavg$
           for $q=0.5, 1, 2, 4, 8, 16$
           for the quantized standard map (red squares)
           vs.\ $1/\ln N$.
           The full lines show $\DqavgCUEN$,
           Eq.~\eqref{eq:D-q-CUE-typical}.
           The grey circles show the corresponding
           random matrix results for one realization of the CUE.
           }
  \label{fig:D-q-std-N-scaling}
\end{figure*}

Figure~\ref{fig:D-q-std-N} shows a comparison
of $\DqN$, computed from one realization of the quantized
standard map for different $N=400$, $2000$, and $10000$,
with the lower-bound $\DqavgCUEN$, Eq.~\eqref{eq:D-q-CUE-typical}.
The agreement with the CUE results of Fig.~\ref{fig:D-q-CUE-N}(a)
is quite good, and improves with increasing $N$.
Thus overall one can conclude from Fig.~\ref{fig:D-q-std-N}
that the multifractal moments of the eigenvectors
of the quantized standard map with fully chaotic dynamics
are very well described by the corresponding random matrix computations.
For the CUE and the quantized standard map,
the $\DqNavg$ show similar deviations from the
analytical prediction~\eqref{eq:D-q-CUE-typical}.

To analyze the scaling of $\DqN$ and $\DqNavg$ towards $1$ in the limit
$N\to\infty$, Fig.~\ref{fig:D-q-std-N-scaling} shows the fractal dimensions
vs.\ $1/\ln N$.  The comparison with $\DqavgCUEN$,
Eq.~\eqref{eq:D-q-CUE-typical}, and the results for the CUE, displayed in
Fig.~\ref{fig:D-q-std-N-scaling}(b), show a similar scaling.  For larger values
of $q$ the fluctuations become more pronounced.  For $\DqN$, displayed in
Fig.~\ref{fig:D-q-std-N-scaling}(a), there is good agreement with the lower
bound provided by $\DqavgCUEN$, Eq.~\eqref{eq:D-q-CUE-typical} when $q$ is
small.  However for larger values of $q$ there are clear deviations of the
fractal dimensions.  With increasing $N$, i.e.\ decreasing $1/\ln N$, the data
approach the lower bound~\eqref{eq:D-q-CUE-typical} from above.  Moreover the
numerical results show that the finite-size corrections both for $\DqN$ and
$\DqNavg$ for the quantized standard map are similar to those for the CUE.

\section{Many body systems}\label{Sec:manybody}

We now turn to another class of systems for which random-matrix theory is often
applied: many-body systems which are neither integrable nor
many-body-localized.  We will present results for a specific spin chain
(the XXZ chain with nearest-neighbor interactions) for two different choices of
parameters.  In addition we have performed similar calculations for other
many-body lattice Hamiltonians, and found the overall multifractality
properties to be very similar.  We thus believe the results presented here to
be qualitatively generic.

\subsubsection{Hamiltonian}

We consider a disorder-free XXZ Heisenberg chain, consisting of $L$ sites and
one spin-1/2 particle on each site, with both nearest-neighbor (NN) and
next-nearest-neighbor (NNN) interactions:
\begin{multline}
H ~=~ J_1 \sum_{i=1}^{L-1}  \left( S_i^+S_{i+1}^- +S_i^-S_{i+1}^+
                                + \Delta_1 S_i^zS_{i+1}^z \right)
\\
~+~ J_2 \sum_{i=2}^{L-2}  \left( S_i^+S_{i+2}^- +S_i^-S_{i+2}^+
                               + \Delta_2 S_i^zS_{i+2}^z \right)
.
\label{eq:H_XXZNNN}
\end{multline}
Here $S_i^\pm = S_i^x \pm \ui S_i^y $ with
$S_i^x = \frac{\hbar}{2} \sigma_i^x$, $S_i^y = \frac{\hbar}{2} \sigma_i^y$, and
$S_i^z = \frac{\hbar}{2} \sigma_i^z$, using the Pauli-matrices acting only on
the $i$-th site.  The summations in~\eqref{eq:H_XXZNNN} are over the site
index.
The XXZ chain with NNN interactions is a canonical example of a non-integrable
many-body system.  As such, the mid-spectrum eigenstates and the dynamics of
this model and its variants have been studied from several perspectives in
recent years (see, e.g, \cite{San2009, SanBorIzr2012b, SteHerPre2013,
  PogWis2016, TorKarTavSan2016, BeuBaeMoeHaq2018,KhaHaqCla2019,
  BorIzrSan2019}).  Of course, the equilibrium (low-energy) properties of such
models have been considered extensively, already in earlier decades, but these
are less relevant to the present work.

In order to avoid reflection symmetry, we have omitted the NNN coupling between
sites $1$ and $3$ (the summation starts from $i=2$ instead of $i=1$).  The NNN
coupling breaks integrability; to keep away from an integrable point we use
$J_2=J_1$.  We also set both couplings $J_{1,2}$ to unity, i.e., energies are
measured in units of $J_1$.

The XXZ chain~\eqref{eq:H_XXZNNN} conserves the total
$S^z = \sum_{i=1}^{L-1} S_i^z$, or equivalently, the number of up-spins or
``particle number'' $M$.  For $M$ up-spins in $L$ sites, the Hilbert space
dimension is $N=\binom{L}{M}$.  As parameters we use
$(\Delta_1,\Delta_2)=(2.0,0.0)$ throughout the text, apart from
Fig.~\ref{fig:D-q-XXZ-N-both}, where in addition
$(\Delta_1,\Delta_2)=(0.8,0.8)$ is used.

We have checked that the system shows the correct GOE level spacing statistics
for either of these parameter sets, e.g., the average $\langle r \rangle$ of
the ratio of successive consecutive-neighbor level spacings~\cite{OgaHus2007,
  AtaBogGirRou2013} is near the value ($\approx0.53$) expected for the GOE. The
ratio of the spacings between two closest
levels~\cite{SriLakTomBae2019} is also near the GOE value ($\approx0.57$).

\subsubsection{Overview: various parts of the spectrum}

\begin{figure}[t]
  \includegraphics{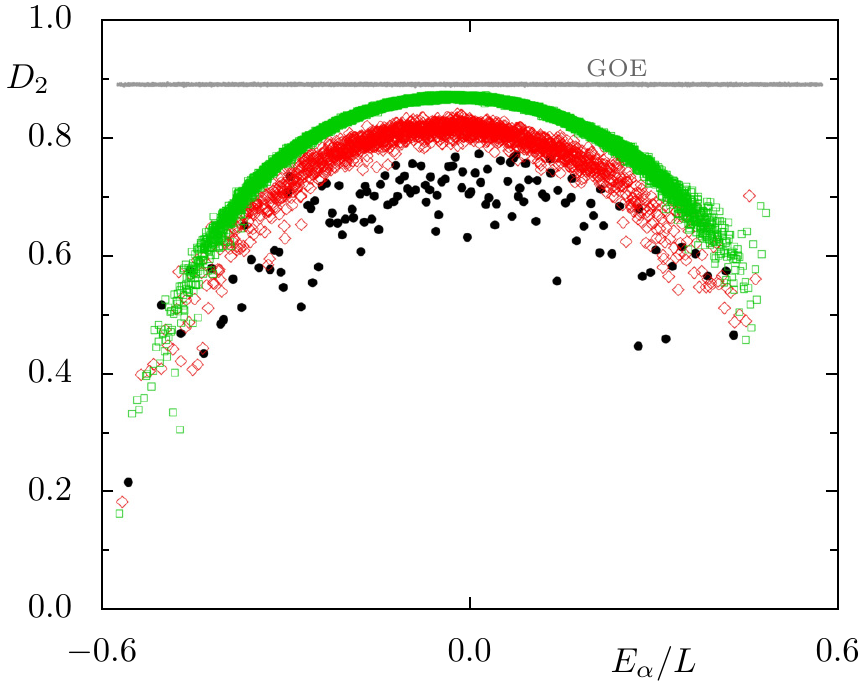}
  \caption{Fractal dimensions $\DqjN$ for $q=2$
           vs.\ the scaled eigenenergies, $E_\alpha/L$,
           for the XXZ spin-chain.
           for $(L, \Nup) = (9, 4)$, $(13, 6)$, $(17, 8)$
           (black circles, red diamonds, green squares).
           The corresponding Hilbert space dimensions
           are $N= 126$, $1716$, $24310$.
           For one realization of the GOE with $N=24310$
           the fractal dimensions $D_2$ are shown as small grey dots
           versus the energies (rescaled to approximately the
           same bandwidth as the spin-chain).
           }
  \label{fig:D-q-XXZ-vs-E}
\end{figure}

We first consider the fractal dimensions of all eigenstates for the XXZ spin
chain~\eqref{eq:H_XXZNNN} in the ergodic regime.
Eigenstates at the very low-energy and very high-energy edges of the spectrum
are multifractal.  Indeed, for the lowest and the highest eigenstates
$
\Dqinf\neq1$ for $q\neq0$~\cite{AtaBog2012,BeuBaeMoeHaq2018}.
For non-integrable systems, it is widely expected that the eigenstates in the
middle of the many-body spectrum behave at least like random-matrix
eigenstates; in fact this expectation may be considered the basic idea behind
the eigenstate thermalization hypothesis~\cite{Deu1991, Sre1994, RigDunOls2008,
  PolSenSilVen2011, AleKafPolRig2016,BorIzrSanZel2016, NeuMar2012,
  BeuMoeHaq2014, Rei2015, NatPor2018, KhaHaqCla2019}.
Thus, we expect that the middle of the spectrum is at least weakly ergodic in
the sense that the corresponding wavefunctions occupy a finite fraction of the
Hilbert space and, thus, $\DqN$ approaches $1$ in the $N\to\infty$ limit.

Figure~\ref{fig:D-q-XXZ-vs-E} illustrates the $N$-dependence of the fractal
dimensions by plotting $D_2(j, N)$ for every eigenstate $j$ of the
non-integrable spin-chain for different values of the Hilbert space dimension
$N$: As the system size increases, the $\DqjN$ values for mid-spectrum
eigenstates move up towards $1$, i.e.\ the eigenstates show the expected
ergodic behavior.  (We will later show that the approach to $1$ is
logarithmically slow.)  In contrast, for the bottom or top of the spectrum
there is no trend towards $1$ which is consistent with the picture that these
eigenstates are multifractal.

\begin{figure}[t]
  \includegraphics{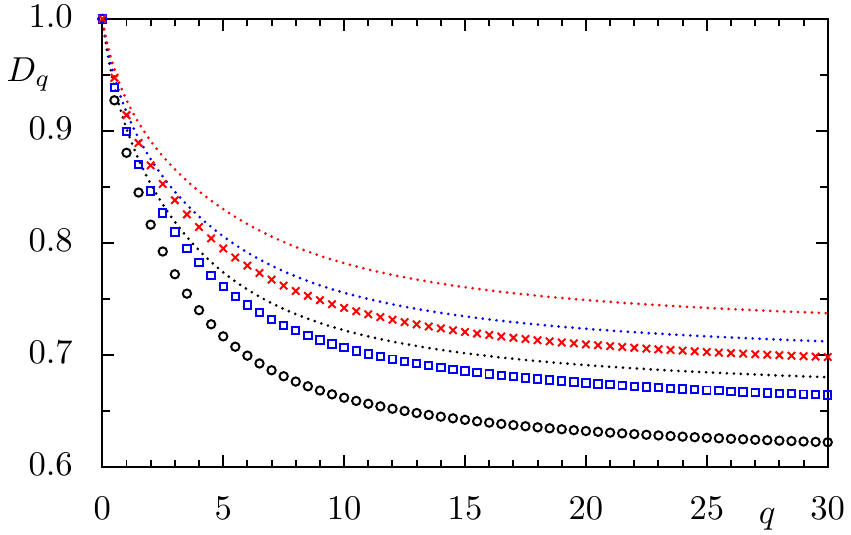}

  \caption{Fractal dimensions $\DqN$ versus moments $q$
           for the XXZ spin chain for different system sizes
           $(L, \Nup) = (13, 6)$, $(15, 7)$, $(17, 8)$
           (black circles, blue squares, red crosses)
           using $250$ states in the middle of the spectrum.
           For comparison numerical results for realizations of the
           GOE results are shown
           as dotted curves of the correspondent colors for
           $N=1716, 6435, 24310$.
           }
  \label{fig:D-q-XXZ-N}
\end{figure}

In contrast, the results for a realization of the GOE show no dependence on the
energy, e.g.\ there is no multifractality near the edges of the spectrum, even
though the spectral density of the GOE does depend on the energy.  This is
shown as the grey points forming a straight line in
Fig.~\ref{fig:D-q-XXZ-vs-E}.  For the purposes of the eigenvector statistics
the results for the COE obtained in Sec.~\ref{sec:COE} are identical to those
of the GOE as only the normalization condition~\eqref{eq:COE-normalization} is
relevant.  Also note that the results for the spin-chain (green squares) with
the same dimension $N=24310$ are well below the GOE result, even in the middle
of the spectrum.  Therefore an important question, to be addressed in the next
section, is how $\DqN$ approaches $1$ for mid-spectrum many-body eigenstates.

\subsubsection{Comparison of the
fractal dimensions with the GOE}

\begin{figure}[b]
  \includegraphics{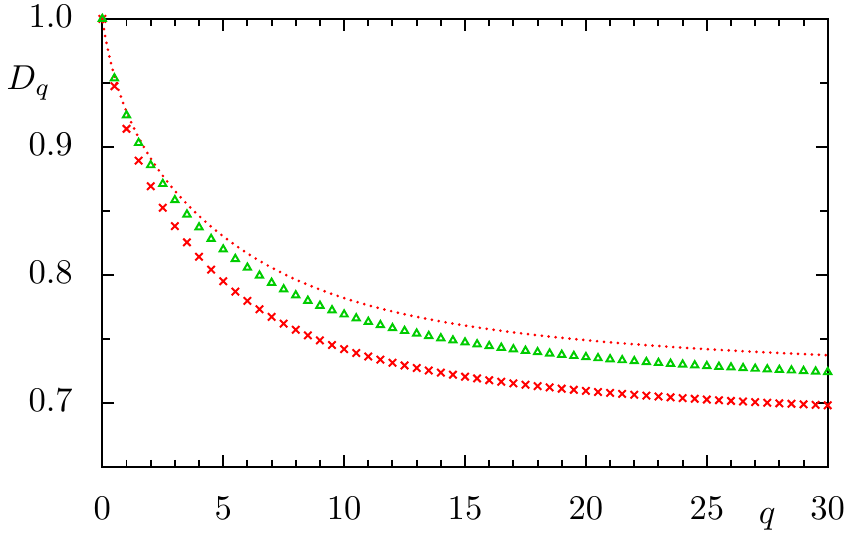}

  \caption{Fractal dimensions $\DqN$ versus $q$
           for the XXZ spin chain for
           $(L, \Nup) = (17, 8)$, such that $N=24310$,
           for $(\Delta_1, \Delta_2)=(2.0, 0.0)$,
           red crosses, as in Fig.~\ref{fig:D-q-XXZ-N},
           and $(\Delta_1, \Delta_2)=(0.8, 0.8)$, green triangles.
           In both cases $250$ states in the middle of the spectrum
           are used.
           The red dotted line is the numerical result
           for one realization of the GOE for the corresponding matrix
           size $N=24310$, as in Fig.~\ref{fig:D-q-XXZ-N}.
           }
  \label{fig:D-q-XXZ-N-both}
\end{figure}

Figure~\ref{fig:D-q-XXZ-N} shows the $q$-dependence of the fractal dimensions
$\DqN$ in the many-body system in comparison with numerical results for the GOE
of the corresponding sizes $N$.  To avoid fluctuations due to finite statistics
we only show the typical fractal dimensions $\Dq$.
The overall shape and size-dependence is qualitatively similar to that in the
COE, CUE, and standard map cases studied in previous sections.  However, the
departure from the random-matrix data is now much stronger: the deviations are
already significant for the smallest moments and become more prominent with
increasing $q$.  With increasing system size $N$ the fractal dimensions $\DqN$
become larger, moving towards $1$, and the difference to the random matrix
results becomes smaller.

The amount of the deviations from the COE prediction are highly system
and parameter specific.  Indeed, even considering the same many-body
model~\eqref{eq:H_XXZNNN} for either $(\Delta_1, \Delta_2)=(2.0, 0.0)$, as
before, or $(\Delta_1,\Delta_2)=(0.8,0.8)$, reveals very different departures
from the GOE results as illustrated in Fig.~\ref{fig:D-q-XXZ-N-both}.

We have also examined the distribution $P(\Dq)$ for the many-body eigenstates
(not shown).  The distribution is qualitatively similar to the one for the COE
or CUE (which is shown in Fig.~\ref{fig:D-q-CUE-distrib}.)

\subsubsection{Size dependence: weak ergodicity}

\begin{figure}
  \includegraphics{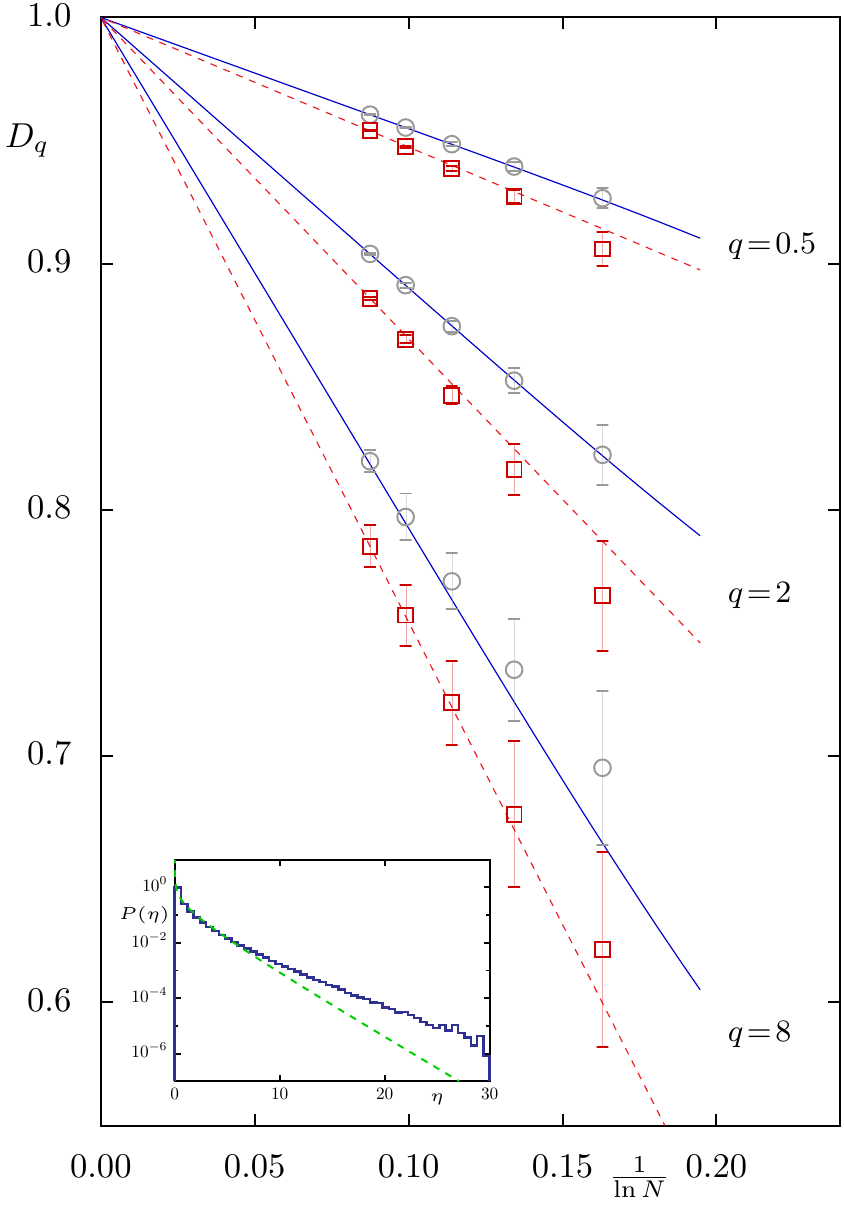}
  \caption{Fractal dimensions $\DqN$ for $q=0.5, 2, 8$ for the XXZ spin-chain
    (red squares) versus $1/\ln N$ for the sequence of system sizes
    $(L,\Nup) = (11,5)$, $(13,6)$, $(15,7)$, $(17, 8)$, $(19, 9)$.  The $250$
    states in the middle of the spectrum are used for all sizes except the
    largest one (for $(L,\Nup) = (19, 9)$ $150$ states are used).  The solid
    blue lines show the COE analytical prediction $\DqavgCOEN$,
    Eq.~\eqref{eq:D-q-COE-typical}.  The grey circles show the corresponding
    random matrix results for one realization of the GOE for $N=462$, $1716$,
    $6435$, $24310$, $92378$.
    The vertical bars indicate the standard
    deviation of the data.  The red dashed lines for each $q$ are guides to the
    eye connecting $(1/\ln N, \DqN) = (0, 1)$ with the value of $\DqN$ at the
    largest available system size (data point with smallest value of
    $1/\ln N$).
    The inset shows the coefficient distribution $P(\eta)$ of the
    250 eigenvectors for $(L, \Nup) = (17, 8)$
    in a semi-logarithmic representation.
    Clear deviations from the Porter-Thomas distribution
    \eqref{eq:distribution-N-limit-COE} of the GOE (green dashed line)
    are found.}
  \label{fig:D-q-xxz-N-scaling}
\end{figure}

To systematically address the scaling limit of the fractal dimensions in the
many-body system Fig.~\ref{fig:D-q-xxz-N-scaling} shows $\DqN$ versus the
$1/\ln N$ (red squares), i.e.\ the inverse logarithm of the Hilbert space
dimension, together with GOE data (grey circles) and the analytical prediction
$\DqavgCOEN$, Eq.~\eqref{eq:D-q-COE-typical}, (solid blue lines) for the
corresponding matrix size $N$ for several moments $q$.  The red dashed straight
lines are guides to the eye connecting $\Dq = 1$ at $N\to\infty$ with the
many-body data point at the largest considered system size. The error bars are
given by the standard deviation of the distribution of the corresponding
fractal dimensions $\DqjN$.

It is clearly seen in Fig.~\ref{fig:D-q-xxz-N-scaling} that the many-body data
approaches $\DqN\to 1$ at $N\to\infty$, however the path of this approach is
different from the one of the GOE: The fractal dimensions $\DqN$ are
smaller, while the standard deviations (shown as error bars) are larger.  This
clearly suggests that the eigenstates of a typical non-integrable many-body
system are only weakly ergodic, i.e., that they only occupy a finite fraction
$\rho$ of the whole Hilbert space.

Following Eq.~\eqref{eq:IqEqcqNDq}, we can express the weak ergodicity
in terms of the scaling of the typical moments
by comparing with the GOE result, Eq.~\eqref{eq:c-q-COE},
\begin{equation}
  \la I_q(j, N) \ra_{\text{typ}}
    = c_q N^{-\Dqinf (q-1)}
    = c_q^{\text{GOE}} N_{\text{eff}}^{-\Dqinf (q-1)} \ .
\end{equation}
where $N_{\text{eff}} = \rho N$ with
fraction $\rho \equiv N_{\text{eff}}/N = (c_q^{\text{GOE}}/c_q)^{1/(\Dqinf(q-1))}<1$.
This shows that the deviation of $c_q$ compared to $c_q^{\text{GOE}}$
corresponds to the effectively reduced fraction of the whole
Hilbert space occupied by weakly-ergodic eigenstates,
compared to ergodic ones of the GOE.

The weak ergodicity also suggests that a standard random matrix ensemble like
the GOE of GUE is not a fully correct description of the statistical properties
of the many-body states, even in the middle of the spectrum.
This can also be seen by examining the coefficient distribution.  In the inset
of Fig.~\ref{fig:D-q-xxz-N-scaling}, the distribution $P(\eta)$ of the
(rescaled) eigenvector components $\eta=N |c_i^{(j)}|^2$ is shown.  There are
clear deviations from the Porter-Thomas distribution
\eqref{eq:distribution-N-limit-COE} of the GOE.  The deviations at the tail of
the distribution are highlighted here by using a logarithmic scale.  Some
deviations from the random-matrix expectation was also noted in
Ref.~\cite{BeuBaeMoeHaq2018}.

\section{Summary and outlook}\label{Sec:summary}

In this work we have addressed the deviations of the eigenstate statistics from
the fully ergodic result --- for random matrix ensembles, a single-particle
system with chaotic classical dynamics, and chaotic many-body systems.  We
analyzed the scaling behavior of the fractal dimensions $\DqN$ which should
approach one in the limit of large system size $N$ if the system is fully
ergodic.

For the standard random matrix ensembles (COE and CUE) we provide analytical
results for the means $\DqNavg$ over individual eigenstates.  This provides a
lower bound for the typical $\DqN$ (logarithmic) averages of eigenstate
moments.  We show that individual realizations of COE and CUE typically match
the predictions only for small $q$, and deviate at larger $q$ due to finite
statistics.  We have provided an estimate of
the value $q^*(N)$ beyond which finite-statistics effects become important:
$q^*(N)$ scales logarithmically with $N$ such that
obtaining agreement at larger $q$ would require averaging over an
exponentially large number of realizations.

For the quantized standard map with classically chaotic dynamics,
the numerical results agree well with
those for realizations of random matrices.  For both random matrices and the
quantized standard map, the approach $\DqN \to 1$ with increasing system size
is slow, and closely follows the form $\Dqinf - f_q/\ln{N}$ for small $q$.  For
larger $q$, there are strong deviations from this form and the data even shows
some curvature when $\DqN$ is plotted against $1/\ln{N}$.  This curvature
implies an $N$-dependence of the quantity $f_q$, or
equivalently, of the quantity
$c_q=\ue^{(q-1)f_q}$ used in Eq.~\eqref{eq:IqEqcqNDq}.

In contrast, the results for the
many-body systems deviate quite significantly from the COE data.  We have
analyzed these deviations in $\DqN$ for different eigenstates, different values
of $q$, and system sizes.  The fractal dimensions of the non-integrable
many-body systems still approach the ergodic limit $\DqN = 1$ in the
thermodynamic limit $N\to \infty$.  However, the path of this approach differs
from the random-matrix one and is system-specific: writing the $N$-dependence
as $\Dqinf - f_q/\ln{N}$ requires $c_q$ to be larger than the GOE value
$c_q^{\text{GOE}}$.  We thus conclude that mid-spectrum many-body eigenstates
are of weakly ergodic nature and occupy only a finite fraction of the whole
Hilbert space.  We speculate that this may result from the fact that
mid-spectrum eigenstates are forced to be orthogonal to the eigenstates at the
spectral edges, which are very special (multifractal).

The present work opens up various new questions.
(1)~The curvature in the
$\DqN$ versus $1/\ln{N}$ plots points to finite-size structures in
random-matrix eigenstates which deserve further study.  If we write $\DqN$ as
$\Dqinf - f_q/\ln{N}$, then $c_q=\ue^{(q-1)f_q}$ is weakly $N$-dependent; the
form of dependence is a non-trivial characterization of finite-size
random-matrix eigenstates which would be interesting to investigate.
(2)~We have
characterized multifractality properties of the quantized standard map at large
$K$, for which the classical counterpart is strongly chaotic.
As $K$ is decreased, the classical dynamics shows a mixed phase space
in which regular motion and chaotic motion coexist on arbitrarily fine
scales. This will change the behavior of $\DqN$ and could lead to
weak ergodicity or multifractality.
(3)~We have only examined many-body models which are nominally chaotic.
The $\DqN$ behaviors of mid-spectrum eigenstates of \emph{integrable} many-body
systems remains an open issue.
(4)~Our results suggest that many-body eigenstates are only weakly ergodic.
This implies that the standard random matrix classes (GOE, GUE) may not be the
optimal random-matrix models for describing the mid-spectrum eigenstates.
Other random matrix classes, such as the power-law-banded random matrices
\cite{MirFyoDitQueSel1996, KraMut1997,MirEve2000,VarBra2000,
  EveMir2008,BogSie2018b}, might be fruitful to examine as models of
eigenstates of non-integrable many-body Hamiltonians.

~  

\acknowledgments

We thank Wouter Beugeling and Paul McClarty for useful discussions.  I.~M.~K.\
acknowledges the support of German Research Foundation (DFG) Grant No.\
KH~425/1-1 and the Russian Foundation for Basic Research Grant No.\
17-52-12044.

\end{document}